\documentclass{article}

\usepackage[english]{babel}

\usepackage[letterpaper,top=2cm,bottom=2cm,left=3cm,right=3cm,marginparwidth=1.75cm]{geometry}
\usepackage{pdflscape}
\usepackage{amssymb}
\usepackage{siunitx}
\PassOptionsToPackage{hyphens}{url}
\usepackage[utf8]{inputenc}
\usepackage{csquotes}
\usepackage{booktabs}
\usepackage{longtable}
\usepackage{adjustbox}
\usepackage{array}
\usepackage{url}
\usepackage{titlesec}
\usepackage{authblk}
\usepackage{xcolor} 

\hypersetup{colorlinks,allcolors=black}
\usepackage{graphicx}
\usepackage{float}
\usepackage{natbib}
\titleformat{\subsection}
  {\mdseries\itshape\large} 
  {\thesubsection}{1em}{} 

\author[1]{Eleanor Wiske Dillon}
\author[1]{Sonia Jaffe}
\author[1]{Sida Peng}
\author[1]{Alexia Cambon}

\affil[1]{Microsoft}

\title{Early Impacts of M365 Copilot\thanks{Corresponding author: eldillon@microsoft.com. We thank the Microsoft Customer Research Program, Modern Work Marketing, and company partners for help carrying out this experiment, Abigail Atchison, Roman Basko, and Fabio Vera for superb data science support, Jack Cenatempo, Esther Plotnick, and Will Wang for excellent research assistance, and Nicole Immorlica and Danielle Li for helpful comments.}}

\begin{document}
\date{January 6, 2025\\ \vspace{15pt}}
\maketitle

\begin{abstract}
Advances in generative AI have rapidly expanded the potential of computers to perform or assist in a wide array of tasks traditionally performed by humans. We analyze a large, real-world randomized experiment of over 6,000 workers at 56 firms to  present some of the earliest evidence on how these technologies are changing the way knowledge workers do their jobs. 
We find substantial time savings on common core tasks across a wide range of industries and occupations: workers who make use of this technology spent half an hour less reading email each week and completed documents 12\% faster.
Despite the newness of the technology, nearly 40\% of workers who were given access to the tool used it regularly in their work throughout the 6-month study. 

\end{abstract}

\section{Introduction}
\label{sec:introduction}

New generative AI tools have been developing rapidly and are now widely used, including by workers in doing their jobs. Microsoft worked with firms across industries to run a large field experiment to measure how access to one of these tools changes work patterns. 
The experiment ran during the early rollout of Microsoft's M365 Copilot (Copilot), a multi-part generative AI tool that integrates generative AI into components of Microsoft's M365 suite (including Word, PowerPoint, Outlook, and Teams). M365 Copilot is designed as a general purpose tool to help workers digest information by summarizing emails, meetings, or documents, create new content by drafting emails, documents, or slide decks, and retrieve information either from the web or across any company email, chat, or document to which the worker has access. 

We worked with firms to randomize access to Copilot and got permission to use several months of anonymized metadata on workers' behaviors in Outlook, Teams, and Office, both before and after access to Copilot. From that data we create worker-week level metrics of activity around sending and reading emails, attending meetings, and creating, reading, and editing documents. The experiment involved over 6,000 workers across multiple industries, occupations, and levels of seniority. All participating workers already spent significant portions of their workday interacting with various M365 products. While most workers also had the option to access other generative AI tools such as ChatGPT, we expect the integration of generative AI into these commonly used work programs may be particularly appealing for these workers.

Copilot is a novel technology, and firms varied in the ways they trained workers to use these new tools or encouraged use. We therefore estimate both intent-to-treat and treatment-on-the-treated effects. The first effect measures the average impact of having access to Copilot, including the behavior of workers who had access but did not use it regularly. The second captures the impact of Copilot for the early adopters who integrated this generative AI tool into their weekly work, relative to workers who look similar to these regular users but did not have access to Copilot because of the experimental design. We expect the long-run impacts of tools like Copilot may fall somewhere between these two estimates; more workers will begin using generative AI as the tools become more familiar and their effective use cases become clearer, but the average benefits may not match the experience of the early adopters.

Over this wide range of workers, we find consistent evidence that Copilot helped workers complete some core tasks in less time. The average worker in our sample spends 3 hours each week reading emails. Workers with access to Copilot reduced this time by 7\% on average (12 minutes per week), through a combination of opening fewer emails and spending less time reading each one. Treated workers who used Copilot at least once a week reduced their weekly time reading emails by half an hour (18\%). We do not see evidence that this time savings caused workers to ignore work; workers with Copilot replied to the same number of email conversations as their colleagues and sent those replies somewhat faster. We find even larger time savings in document creation. Workers with access to Copilot create modestly more Word documents and read others' documents more frequently. When they are the primary editor on a document, they complete it half a day faster on average, for those who used Copilot regularly, the effect was almost a whole day. (The average document takes 7 days to complete).

Workers are most likely to use Copilot to summarize meetings, but we do not see similar reductions in time spent in meetings. We hypothesize that the ability of generative AI to transcribe and summarize meetings can make live meetings more effective, but workers are as likely to respond to that improvement by shifting more work into meetings, for example holding more brainstorming sessions, as they are to use that efficiency to clear their schedule. The lack of effect on meeting time could also be due to the fact that workers' colleagues mostly did not have access to Copilot, so the impact might be different in the case of more widespread adoption. Looking at the variation across firms, we see as many firms with significant decreases in meeting times for their workers with Copilot as firms with significant increases.

This work joins a small set of early studies of generative AI in real workplaces. \cite{brynjolfsson2023generative} found that an AI-driven conversational assistant allowed customer chat support agents to resolve 14\% more issues per hour. \cite{cui2024impact} found that software developers using GitHub Copilot completed 26\% more tasks. In a study of Kenyan entrepreneurs, \cite{otis2023uneven} found that a generative AI-powered support tool improved an index of entrepreneurial performance by 0.19 standard deviations, but only among entrepreneurs with above-median performance before gaining access to the tool. The magnitude of impacts we estimate align with these other studies, albeit in very different contexts. 

These real-world studies complement a larger body of work understanding the impact of generative AI on controlled tasks in a lab setting.\footnote{See, for example, \cite{peng2023impact}; \cite{Spatharioti2024search}; \cite{vaithilingam2022expectation}; \cite{campero2022}; \cite{noy2023experimental}; \cite{edelman2024impact}; \cite{edelman2024impact2}, \cite{dell2023navigating}, \cite{cambon2023early}} Unlike our experiment, these lab studies observe all dimensions of participants' output and can measure both time savings and changes in productivity or quality. For example, \cite{noy2023experimental} found participants using ChatGPT to complete professional writing tasks demonstrated decreases of 0.8 standard deviations in time taken and improvements of 0.4 standard deviations in output quality. However, these studies where participants are explicitly instructed to use generative AI tools to complete a single task, generally chosen to be particularly suited to the strengths of generative AI, can provide only limited insights into real-world productivity gains. Most jobs contain a mix of tasks that may or may not benefit from the use of generative AI and workers may not use generative AI tools for all tasks where they might experience gains.

Generative AI tools are being adopted by workers at an unprecedented rate. In two recent surveys, \cite{humlum2024adoption} find 27\% of Danish workers in exposed occupations used ChatGPT at work by the end of 2023, only one year after its release, and \cite{blandin2024rapid} find 24\% of a representative sample of U.S. workers used generative AI at work by August 2024. These estimates are broadly in line with our findings, but suggest much faster adoption than earlier innovative technologies. For example, \cite{acemoglu2022automation} find that only 45\% of manufacturing workers in the U.S. were exposed to robotics technology in 2016-18, 50 years after the first major advances in that space.\footnote{Looking at computers, 25\% of workers used them in 1984, fourteen years after the Intel microprocessor. More recently, Geoffrey Hinton's team won the ImageNet competition with a system based on neural nets in 2012 \cite{thompson2021robotics}, but \cite{acemoglu2022automation} find that only about 30\% of U.S. workers in the information sector (the most exposed group) were using any kind of AI technology at work in 2016-18.} This study provides important early insights into how this rapid technological adoption may begin to reshape work.

\section{Experiment Design and Implementation}
\label{sec:Expdesign}

During the early rollout of M365 Copilot, firms could purchase only a small number of licenses. We worked with 56 firms to randomize the allocation of some of these scarce licenses within a pool of suitable workers, over 6,000 in total. Firms maintained the random assignment for at least six months, during which we used product telemetry data that Microsoft already collects to track worker actions around sending and reading emails, attending meetings, and creating documents. To maintain workers' and firms' privacy, we do not observe anything about the content produced by these workers, so our focus is on how workers allocate their time and interact with each other rather than on the quality of their work; moreover, we only analyze aggregate patterns, not individual activity.  Only very large, mainly multi-national firms were invited to participate the experiment. The experiment includes workers from many countries, with the majority based in the United States or Europe. Table \ref{tab:industries} illustrates the wide range of industries represented in the study.

\begin{table}[H]
\centering
\begin{tabular}{|l|c|}
\hline
\textbf{Industry} & \textbf{Firms} \\ \hline
Telecommunications & 9 \\ \hline
Professional Services & 8 \\ \hline
Banking and Financial Services & 7 \\ \hline
Construction and Manufacturing & 6 \\ \hline
Energy and Mining & 5 \\ \hline
Consumer Goods & 4 \\ \hline
Consumer Services & 4 \\ \hline
Retail & 4 \\ \hline
Government and Public Services & 3 \\ \hline
Insurance & 3 \\ \hline
Technology and Software & 3 \\ \hline
\end{tabular}
\caption{Industry Mix of Participating Firms}
\label{tab:industries}
\end{table}

\begin{figure}[H]
    \centering
    \includegraphics[width=0.45\textwidth]{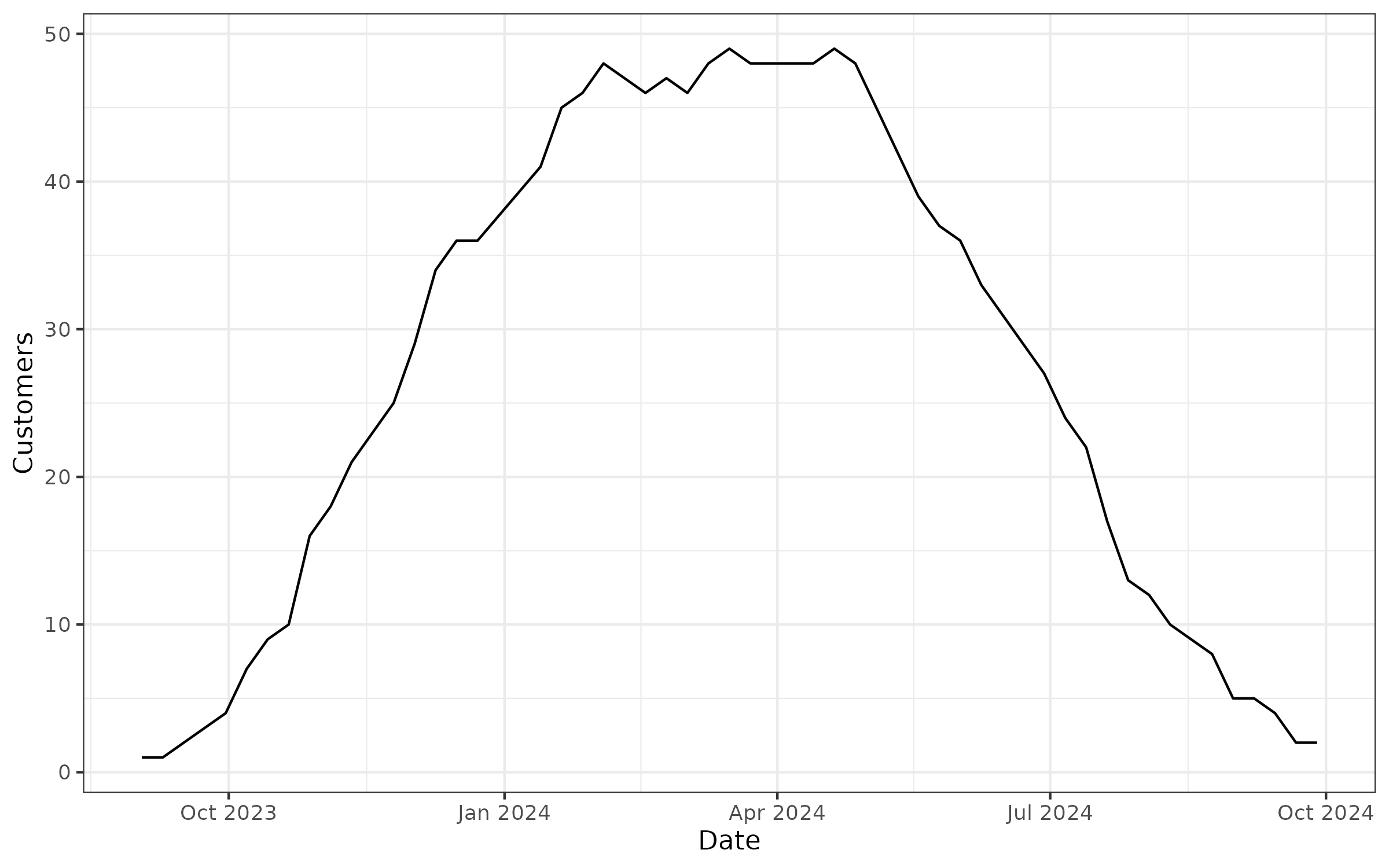}
    \caption{Firms Participating In The Experiment Over Time}
    \label{fig:timeline}
\end{figure}

The study spanned one year, during which firms moved in and out of the sample. Most participating firms maintained the randomized allocation for the full six months. Figure \ref{fig:timeline} illustrates the number of partipating firms over the study window. During the experiment, we collected and analyzed millions of M365 usage signals for both the treated and designated control workers. These data pipelines were established to track key metrics in email, document editing, and meeting workflows. We also tracked adherence to the random assignment. This design ensured robust comparisons between the treated and control groups, enabling us to assess the direct effects of Copilot on workplace behaviors.

\begin{figure}[H]
\label{fig:usage}
    \centering
    \begin{minipage}{0.45\textwidth}
        \centering
        \includegraphics[width=\textwidth]{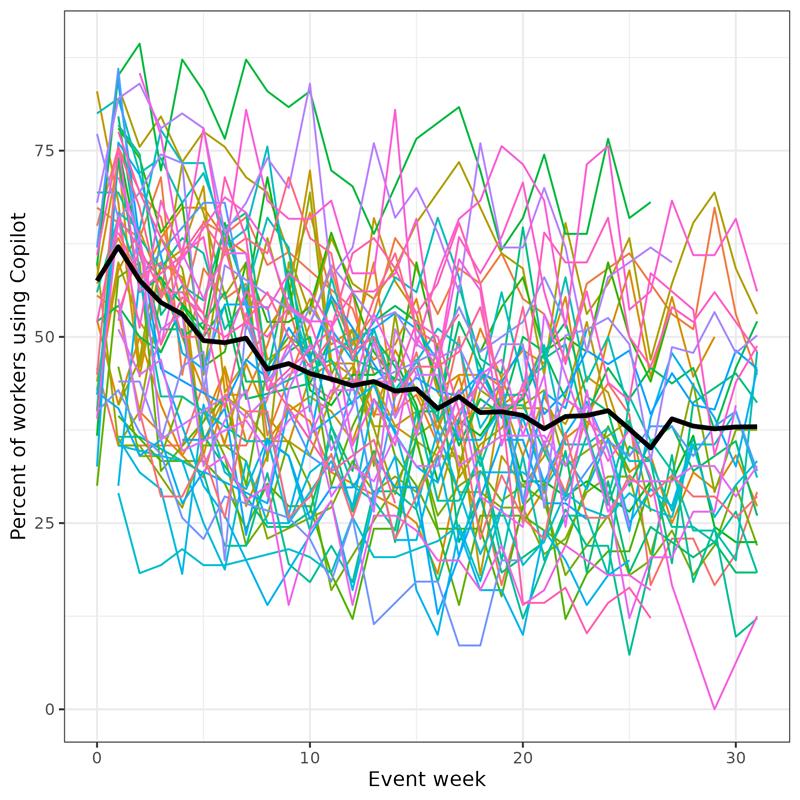}
        \caption{Active Copilot Usage – All Firms and Median \label{fig:firm_usage}}
    \end{minipage}
    \hfill
    \begin{minipage}{0.45\textwidth}
        \centering
        \includegraphics[width=\textwidth]{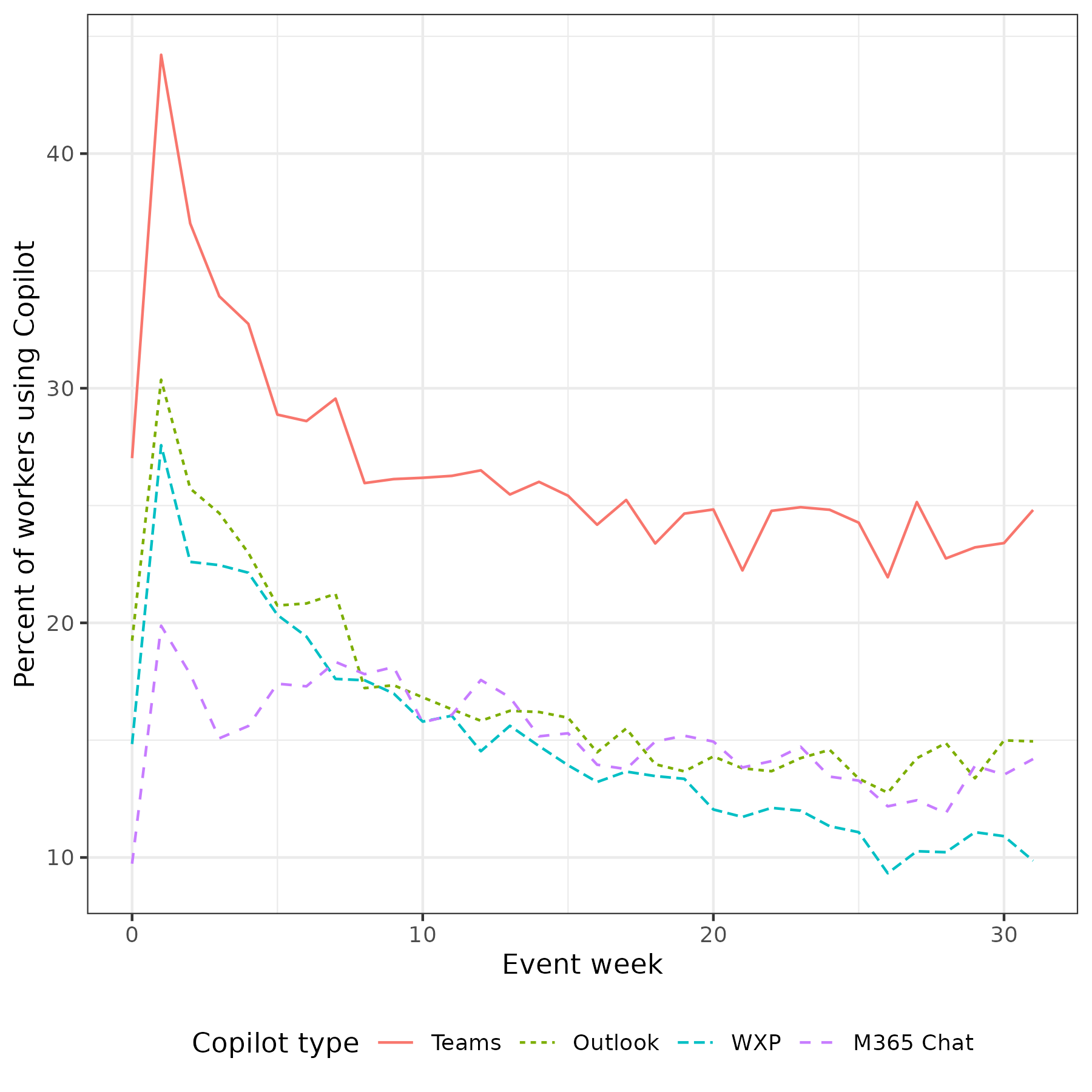}
        \caption{Copilot Usage by App - Across Firms\label{fig:app_usage}}
    \end{minipage}
\end{figure}

On average, 38\% of workers in the experiment who were randomly allocated a Copilot license used Copilot for work each week. Figure \ref{fig:firm_usage} shows how this varied over time and across firms, with average usage at some firms ranging from as high as 75\% to as low as 9\%. These usage rates reflect only engagement with M365 Copilot, not with generative AI tools overall. Firms varied in how they identified workers for the experiment, but most did not put their highest priority workers into this study because all participants had only a 50\% chance of receiving a license, due to the random assignment. Figure \ref{fig:app_usage} shows usage of Copilot by app; on average, the Teams Copilot was the most frequently used, with Outlook, the  M365 Copilot chat interface, and Word, Excel, and PowerPoint combined all seeing about equal usage.

\section{Methods}
\label{sec:Expdesign2}

Our data comprises weekly outcomes in various Microsoft applications for each worker. Firms entered the experiment throughout late 2023 and early 2024, as shown in Figure \ref{fig:timeline}. We collected data starting on May 5, 2023 until firm-specific experimental end dates. In all analyses, we drop outcomes for the weeks ending December 29, 2023 and January 5, 2024 as work activity was low across all workers during this holiday period.

We estimate the effect of receiving a Copilot license on these outcomes using a model with individual, year-month, and firm\textendash event month fixed effects:
\[Y_{it} = \alpha_i + \delta_t + \gamma_{f,\tau_{f}} + \beta Z_{it} + \epsilon_{it}\]
where $f$ is the firm at which worker $i$ works, $\tau_{f}$ is the current event month for firm $f$, and $Z_{it}$ is the worker's experimental assignment. We use Newey-West standard errors with a lag of 9 weeks to handle autocorrelation in outcomes.
This difference-in-differences (DiD) approach uses historical outcome data from before the onset of the experiment to separate persistent differences in workers' behavior due to other factors, such as different roles, from the new impact of generative AI use. This approach leads to more precise estimates than simply comparing means, especially when the sample size is small. 

Organizations may be interested in the effect of giving someone a Copilot license, including the chance that they do not use it. However, the effect of actively using Copilot is also important, both from a research perspective and  for organizations with strong training programs or who plan to reassign licenses that are not used regularly.   In the analyses below, we use an instrumental variable analysis to estimate the effect of a Copilot license on people who actually use Copilot (Copilot users), instead of just those who received a license (licensees). 

A subset of participating firms also shared information on job titles for  workers in the experiment. We hand coded these job titles to separate managers and individual contributors and consider some outcomes separately for these two subgroups.

\section{Results}
We analyze how Copilot affected behaviors across email, documents, and meetings. Analysis broken down by firm industry is in the Appendix.
\subsection{Email}
\label{sec:Email}
The introduction of Copilot has led to significant changes in email-related behaviors among workers. The average worker in the control group  spent 2.8 hours each week reading email and licensees spent 12 fewer minutes reading emails each week, a 7\% decrease. This time savings reflects a combination of fewer individual emails read each week, 9 fewer per week\footnote{In our measure, one email read represents a unique email read at least once by a worker in a calendar week. The same email could contribute to a worker's emails read count in multiple weeks if that worker rereads it.}, and an increase in reading efficiency, as workers with Copilot spend a few seconds less reading each email on average. Copilot users spent 18\% less time reading email, saving more than half an hour each week. We hypothesize that some combination of Outlook Summarize and searching for email content in M365 Copilot Chat allows users to spend less time with individual emails.

\begin{table}[H]
\centering
\begin{tabular}{m{0.3\textwidth}>{\centering\arraybackslash}m{0.15\textwidth}>{\centering\arraybackslash}m{0.2\textwidth}>{\centering\arraybackslash}m{0.22\textwidth}}
\hline
\textbf{Metric} & \textbf{Control Mean} & \textbf{Copilot License Effect (SE)} & \textbf{Copilot Users (SE)} \\
\hline
Messages Read & 164.42 & -9.31** & -23.58** \\
 &  & (0.76) & (1.92) \\
Reply Conversation Length & 3.86 & -0.03** & -0.07* \\
 &  & (0.01) & (0.02) \\
Reply Conversation Count & 13.50 & -0.23* & -0.56* \\
 &  & (0.08) & (0.19) \\
Read Duration (min) & 1.16 & -0.04** & -0.10** \\
 &  & (0.01) & (0.02) \\
log(Read Duration) & --- & -0.06** & -0.15** \\
 &  & (0.0001) & (0.0002) \\
Total Read Duration (min) & 170.16 & -12.30** & -31.16** \\
 &  & (1.23) & (3.11) \\
Time to Reply from  & 476.31 & -18.37** & -46.35** \\
Read Start Time (min) &  & (3.94) & (9.96) \\
\hline
\multicolumn{4}{l}{\footnotesize * $p < 0.05$, ** $p < 0.01$} \\
\end{tabular}
\caption{Effect of Copilot on Email Metrics - Full Panel}
\label{table:email_full_panel}
\end{table}

This time savings does not appear to come at the cost of responsiveness. Workers with Copilot replied to roughly the same number of conversations per week (13.3 vs. 13.5) and had conversations of roughly the same length. Copilot licensees replied to emails 18 minutes (4\%) faster on average, and Copilot users replied nearly 10\% faster (46 minutes). Together, these results suggest that Copilot helps workers effectively triage their email inboxes, rather than facilitating inattention.

\begin{table}[H]
\centering
\begin{tabular}{m{0.4\textwidth} >{\centering\arraybackslash}m{0.075\textwidth} >{\centering\arraybackslash}m{0.075\textwidth} >{\centering\arraybackslash}m{0.15\textwidth} >{\centering\arraybackslash}m{0.15\textwidth}}
\hline
\textbf{Metric} & \multicolumn{2}{c}{\textbf{Control Mean}} & \multicolumn{2}{c}{\textbf{Copilot License Effect (SE)}} \\
 & \textbf{IC} & \textbf{Mgr} & \textbf{IC} & \textbf{Mgr} \\
\hline
Messages Read & 131.36 & 199.21 & -5.02* & -10.36** \\
 &  &  & (2.27) & (3.45) \\
log(Read Duration) & --- & --- & -0.05** & -0.03 \\
 &  &  & (0.02) & (0.02) \\
Total Read Duration (min) & 157.33 & 221.42 & -7.91 & -4.73 \\
 &  &  & (4.04) & (6.16) \\
Time to Reply from  & 434.41 & 446.27 & -44.40** & 6.13 \\
Read Start Time (min) & &  & (13.86) & (14.00) \\
\hline
\multicolumn{5}{l}{\footnotesize * $p < 0.05$, ** $p < 0.01$} \\
\end{tabular}
\caption{Effect of Copilot on Email Metrics - Individual Contributor (IC), Manager (Mgr) Panels}
\label{table:email_ic_mgr_panels}
\end{table}

The impact of Copilot varied by seniority. Both managers and individual contributors (ICs) saved time reading emails overall; however, managers achieved this mainly by reading fewer emails per week while individual contributors mainly saved more time on each email. For these smaller samples, we focus on the effects on log(minutes spent per email), which is less noisy than the raw time spent. A 0.06 decrease in log minutes read can be interpreted as a roughly 6\% decrease. ICs were the primary drivers of the faster reply times, with Copilot users replying to emails 44 minutes faster on average, a 10\% improvement, while managers showed no change in reply time.

\subsection{Documents}
\label{sec:documents}

We expect that the generative capacities of M365 Copilot may help workers produce Word documents more quickly. That speed can have additional ripple effects on how people work. They may have time to contribute to more documents; they may finish rough drafts earlier and take the time to elicit feedback from more colleagues; alternatively, they may co-write less often with colleagues, using Copilot as a substitute writing partner. 

We find small but precisely estimated impacts of Copilot access and use on most of these outcomes. While nearly every worker in our sample uses Outlook regularly to read and write email, not every worker uses Word, Excel or PowerPoint as often. Before the beginning of the experiment, the average worker in our sample made edits to four Word, Excel, or PowerPoint documents in a given week, but was the primary editor of a Word document less than once every other week on average. Our estimated average treatment effects are therefore mostly small in absolute terms, because they include many zero effects from weeks when some workers simply didn't use Word during their work.

\begin{table}[H]
\centering
\begin{tabular}
{m{0.33\textwidth}>{\centering\arraybackslash}m{0.15\textwidth}>{\centering\arraybackslash}m{0.2\textwidth}>{\centering\arraybackslash}m{0.22\textwidth}}
\hline
\textbf{Metric} & \textbf{Control Mean} & \textbf{Copilot License Effect (SE)} & \textbf{Copilot Users (SE)} \\
\hline
Wrd, Exl, Ppt Files Edited & 4.295 & 0.069* & 0.166* \\
 &  & (0.032) & (0.078) \\
Word Docs Created  & 0.438 & 0.020* & 0.495* \\
 &  & (0.008) & (0.019) \\
Primary Editor Word Docs & 0.437 & 0.023** & 0.056** \\
 &  & (0.008) & (0.019) \\
Read Other's Word Docs & 0.526 & 0.031** & 0.075** \\
&  & (0.008) & (0.021) \\
Time to Complete (days) & 7.505 & -0.477** & -0.933**\\
&  & (0.177) & (0.346) \\
Completed Word Docs in $\leq$ 1 wk & 0.314 & 0.012 & 0.029\\
&  & (0.006) & (0.015) \\
Completed Word Docs in $>$ 1 wk  & 0.103 & -0.0002 & -0.0005\\
&  & (0.003) & (0.006) \\
\hline
\multicolumn{4}{l}{\footnotesize * $p < 0.05$, ** $p < 0.01$. Standard errors in parentheses. All metrics are total counts or averages per work week. } \\
\multicolumn{4}{l}{\footnotesize A worker is identified as the creator of a Word document if they make the first edit. They are the primary } \\
\multicolumn{4}{l}{\footnotesize contributor if they make the most edits. Time to complete and completed document counts are calculated }\\
\multicolumn{4}{l}{\footnotesize 
only for documents where the target worker is the primary editor.} \\
\end{tabular}
\caption{Effect of Copilot on Word, Excel, and PowerPoint 
Document Metrics}
\label{table:wxp_odsp_full_panel}
\end{table}

Copilot drives small increases in the overall number if Word, Excel, and PowerPoint files edited, 0.07 more files edited per week on average (a 1.6\% increase from the mean among untreated users) for Copilot licensees  and .17 more documents (3.9\%) for Copilot users. We see larger increases in the number of Word documents that workers initiate (4.6\% and 11.2\% for licensees  and users) and for Word documents where that worker is the primary contributor (5.2\% and 12.8\% more). We find that Copilot licensees (users) read 6\% (14\%) more of their colleagues' Word documents while they are being actively edited, which we interpret as providing feedback. 

We also measure the time workers take to complete Word documents directly, counting the time from when the document is first begun to when 90\% of the total edits to the document have been completed.\footnote{Some documents show stray edits long after most active work is done, which may be accidental. This 90\% cutoff aligns most often with other indicators of completion such as attaching to an email or printing to pdf.} On average, workers with access to Copilot complete Word documents nearly half a day faster on average, a 6\% improvement, and workers who use Copilot complete documents nearly a full day faster (12\%). The final two rows of Table \ref{table:wxp_odsp_full_panel} help us interpret this decrease in completion time. The shorter average completion time is driven mainly by an increase in the number of quickly-completed Word documents, without a decrease in documents that take more than a week to complete. (Remember that these workers are the primary editor on more documents overall.)

\subsection{Meetings}
\label{sec:meetings}
We looked at Copilot’s effect on the number of Teams meetings attended and total time spent in meetings each week. We find that receiving a Copilot license leads to small and statistically insignificant increases in both metrics, but also a (statistically significant) decrease in the proportion of scheduled meeting time during which people actually met, i.e., Copilot licensees are likelier to join meetings later and leave meetings earlier. 

\begin{table}[H]
\centering
\begin{tabular}{m{0.4\textwidth}>{\centering\arraybackslash}m{0.1\textwidth}>{\centering\arraybackslash}m{0.2\textwidth}>{\centering\arraybackslash}m{0.15\textwidth}}
\hline
\textbf{Metric} & \textbf{Control Mean} & \textbf{Copilot License Effect (SE)} & \textbf{Copilot Users (SE)} \\
\hline
Total Session Duration Minutes & 248.67 & 2.38 & 5.78 \\
 & & (1.32) & (3.19) \\
Total Scheduled Meetings Attended & 10.40 & 0.11 & 0.26 \\
 & & (0.05) & (0.13) \\
Joined Scheduled Meeting 5 Min Late & 1.71 & 0.05** & 0.12** \\
 & & (0.01) & (0.03) \\
Left Scheduled Meeting 10 Min Early & 1.72 & 0.04** & 0.11** \\
 & & (0.01) & (0.03) \\
\hline
\multicolumn{4}{l}{\footnotesize * $p < 0.05$, ** $p < 0.01$} \\
\end{tabular}%
\caption{Effect of Copilot on Teams Meeting Metrics}
\label{table:meetings_full_panel}
\end{table}

As shown in Figure \ref{fig:usage}, the Teams Copilot was the most frequently used component of the M365 Copilot suite throughout our sample, suggesting that workers found value in it. We expect that Copilot may have complex and counterbalancing effects on how work meetings operate. First, Copilot may increase meeting efficiency – more can get done in less time, so people can get time back at the end of a meeting and may gradually be able to spend less time in meetings. Pushing in the opposite direction, as Copilot makes meetings more effective, teams may be eager to use it for a wider range of projects or tasks.\footnote{One customer gave an example of how they used to create some documents by taking turns on a document where each person wrote their part, but now they have a meeting to discuss the topic and have Copilot draft the document for them based on the transcript.} Furthermore, meetings previously held offline or on other platforms might be moved onto Teams to take advantage of Copilot’s summary features, leading us to observe more online meetings, even if the total number of meetings remained the same.

\begin{figure}[H]
\label{fig:meetings}
    \centering
    \includegraphics[width=0.7\textwidth]{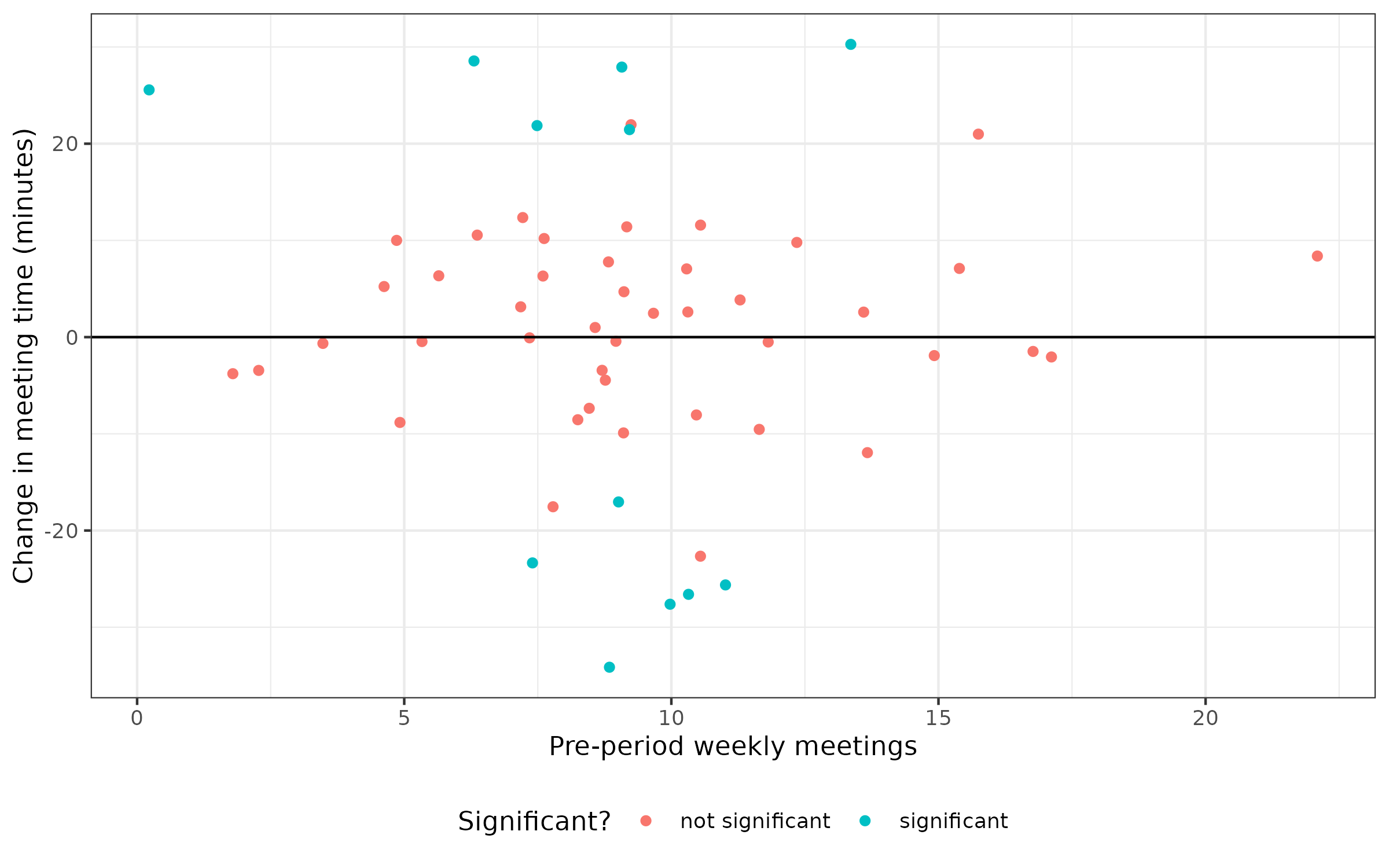}
    \caption{Effect on Total Time in Teams Meetings by Firm Avg. Pre-period Weekly Meetings}
\end{figure}

Figure \ref{fig:meetings} reveals that the negligible average effects we find on weekly meeting duration reflect a combination of large decreases in meeting time at some firms and large increases in others. The sign of the effect on total meeting time is not strongly correlated with companies' use of Teams meetings prior to the experiment, suggesting that other organizational factors determine whether workers react to these new capacities by cutting back on meetings or shifting more work into live collaboration. 

\begin{table}[H]
\centering
\begin{tabular}{m{0.4\textwidth}>{\centering\arraybackslash}m{0.175\textwidth}>{\centering\arraybackslash}m{0.325\textwidth}}
\hline
\textbf{Metric} & \textbf{Control Mean} & \textbf{Copilot License Effect (SE)}  \\
\hline
Total Session Duration Minutes & 200.83 & 26.52** \\
- Positive Firms &  & (3.94) \\
Total Session Duration Minutes & 267.28 & -25.19** \\
- Negative Firms &  &  (4.29) \\
\hline
\multicolumn{3}{l}{\footnotesize * $p < 0.05$, ** $p < 0.01$} \\
\end{tabular}%
\caption{Effect of Copilot on Teams Meetings Metrics - Pooled Positive, Negative Firm Samples}
\label{table:meetings_pooled_panels}
\end{table}

Workers at the six firms that showed statistically significant increases in average meeting time spent 26 minutes more in meeting each week, a 13\% increase from their own pre-experiment mean. Workers at firms that experienced significant decreases overall spent 25 minutes less per week, a 9\% fall.

\section{Conclusion}
\label{sec:conclusion}

This study provides some of the earliest signals of how use of generative AI is beginning to shape the patterns of work for knowledge workers. We observed workers during the first year in which generative AI tools were beginning to gain broad use at work. While their colleagues may have been using other generative AI tools, the workers we studied were often the only members of their team with access to M365 Copilot during the early roll-out period. We therefore expect that the patterns we find are only the beginning of a new era of work. Only as use of these tools become more common, and workers deepen their understanding of their potential, will teams and organizations begin to transform the structure of jobs and work and realize the full potential of these new technologies.



\appendix
\section*{Appendix}
Tables \ref{table:copilot_by_industry} and \ref{table:copilot_by_industry_groups} replicate the analyses above using only firms in a given industry or industry group. 
\section{Results by Industry}
\begin{landscape}
\begin{table}[H]
\centering
\resizebox{\linewidth}{!}{%
\begin{tabular}{lcccccccc}
\hline
\textbf{Metric} 
& \multicolumn{2}{c}{\textbf{Telecommunications}} 
& \multicolumn{2}{c}{\textbf{Professional Services}} 
& \multicolumn{2}{c}{\textbf{Banking and Financial Services}} 
& \multicolumn{2}{c}{\textbf{Construction and Manufacturing}} \\
 & \textbf{Copilot License Effect (SE)} & \textbf{Copilot User (SE)} 
 & \textbf{Copilot License Effect (SE)} & \textbf{Copilot User (SE)} 
 & \textbf{Copilot License Effect (SE)} & \textbf{Copilot User (SE)} 
 & \textbf{Copilot License Effect (SE)} & \textbf{Copilot User (SE)} \\
\hline
Messages Read & -6.04** (1.82) & -13.82** (4.16) & -8.13** (1.80) & -22.53** (4.99) & -9.53** (2.42) & -22.56** (5.72) & -1.18 (2.49) & -4.94 (10.36) \\
Reply Conversation Length & 0.02 (0.03) & 0.04 (0.05) & -0.07* (0.03) & -0.18* (0.07) & -0.04 (0.03) & -0.09 (0.07) & -0.02 (0.03) & -0.07 (0.11) \\
Reply Conversation Count & 0.05 (0.20) & 0.11 (0.45) & -0.18 (0.21) & -0.51 (0.59) & -0.41 (0.22) & -0.93 (0.51) & -0.05 (0.28) & -0.18 (1.07) \\
Read Duration (min) & -0.05** (0.02) & -0.13** (0.04) & -0.010 (0.02) & -0.03 (0.06) & -0.04 (0.02) & -0.09 (0.05) & -0.03 (0.03) & -0.15 (0.11) \\
log(Read Duration) & -0.06** (0.014) & -0.14** (0.03) & -0.04* (0.02) & -0.11* (0.05) & -0.07** (0.02) & -0.16** (0.04) & -0.04 (0.02) & -0.15 (0.08) \\
Total Read Duration (min) & -10.95** (2.84) & -25.07** (6.52) & -6.68* (3.19) & -18.51* (8.81) & -12.55** (4.27) & -29.74** (10.09) & -9.93* (4.39) & -41.45* (18.48) \\
Time to Reply from Read Start Time (min) & -22.48* (9.10) & -51.21* (20.70) & -35.49** (11.23) & -95.24** (30.25) & -29.90* (11.78) & -71.64* (28.24) & -13.03 (13.43) & -53.58 (55.35) \\
Total Session Duration (min) & 7.43 (3.76) & 16.51 (8.32) & -7.89* (3.55) & -21.92* (9.93) & 6.43 (4.03) & 14.71 (9.19) & -0.91 (3.33) & -3.47 (12.74) \\
Total Scheduled Meetings Attended & 0.36* (0.16) & 0.81* (0.35) & -0.24 (0.15) & -0.66 (0.41) & 0.30 (0.17) & 0.68 (0.39) & -0.04 (0.14) & -0.16 (0.52) \\
Joined Scheduled Meeting 5 Min Late & 0.05 (0.04) & 0.10 (0.08) & -0.004 (0.03) & -0.011 (0.08) & 0.08* (0.04) & 0.17* (0.08) & -0.002 (0.04) & -0.009 (0.14) \\
Left Scheduled Meeting 10 Min Early & 0.08 (0.04) & 0.17 (0.09) & 0.007 (0.03) & 0.02 (0.08) & 0.07 (0.04) & 0.15 (0.09) & -0.009 (0.04) & -0.04 (0.14) \\
Wrd, Exl, Ppt Files Edited & 0.14 (0.08) & 0.31 (0.17) & 0.10 (0.10) & 0.29 (0.26) & -0.02 (0.09) & -0.06 (0.20) & 0.005 (0.09) & 0.02 (0.33) \\
Word Docs Created & 0.005 (0.02) & 0.011 (0.04) & 0.02 (0.03) & 0.05 (0.07) & -0.004 (0.02) & -0.008 (0.05) & 0.02 (0.03) & 0.07 (0.11) \\
Primary Editor Word Docs & 0.007 (0.02) & 0.02 (0.04) & 0.008 (0.03) & 0.02 (0.07) & 0.008 (0.02) & 0.02 (0.05) & 0.03 (0.03) & 0.11 (0.11) \\
Read Others' Word Docs & 0.04* (0.02) & 0.09* (0.04) & 0.06 (0.03) & 0.16 (0.08) & -0.02 (0.02) & -0.05 (0.05) & -0.008 (0.03) & -0.03 (0.12) \\
Time to Complete (days) & -1.11* (0.46) & -1.97* (0.82) & 0.13 (0.46) & 0.28 (0.98) & -0.94 (0.50) & -1.64 (0.87) & -0.18 (0.60) & -0.58 (1.92) \\
Completed Word Docs in $\le$ 1 wk & -0.009 (0.014) & -0.02 (0.03) & -0.009 (0.02) & -0.03 (0.05) & 0.02 (0.02) & 0.04 (0.04) & 0.005 (0.02) & 0.02 (0.09) \\
Completed Word Docs in $>$ 1 wk & -0.008 (0.006) & -0.02 (0.013) & 0.003 (0.009) & 0.007 (0.03) & -0.008 (0.007) & -0.02 (0.02) & 0.010 (0.009) & 0.04 (0.04) \\
\hline
\multicolumn{9}{l}{\footnotesize * $p < 0.05$, ** $p < 0.01$} \\
\end{tabular}%
}
\caption{Effect of Copilot on All Metrics by Industry}
\label{table:copilot_by_industry}
\end{table}

\begin{table}[H]
\centering
\resizebox{\linewidth}{!}{%
\begin{tabular}{lcccccc}
\hline
\textbf{Metric} 
& \multicolumn{2}{c}{\textbf{Retail + Consumer Goods}} 
& \multicolumn{2}{c}{\textbf{Construction \& Manufacturing + Energy \& Mining}} 
& \multicolumn{2}{c}{\textbf{Telecommunications + Technology}} \\
 & \textbf{Copilot License Effect (SE)} & \textbf{Copilot User (SE)} 
 & \textbf{Copilot License Effect (SE)} & \textbf{Copilot User (SE)} 
 & \textbf{Copilot License Effect (SE)} & \textbf{Copilot User (SE)} \\
\hline
Messages Read & -17.56** (2.32) & -39.25** (5.15) & -9.98** (1.53) & -27.86** (4.27) & -6.58** (1.69) & -15.37** (3.94) \\
Reply Conversation Length & -0.03 (0.03) & -0.05 (0.06) & -0.03 (0.02) & -0.08 (0.05) & 0.003 (0.02) & 0.007 (0.05) \\
Reply Conversation Count & -0.24 (0.23) & -0.51 (0.49) & -0.012 (0.17) & -0.03 (0.45) & -0.12 (0.20) & -0.27 (0.44) \\
Read Duration (min) & -0.007 (0.02) & -0.02 (0.05) & -0.04** (0.013) & -0.10** (0.04) & -0.06** (0.015) & -0.15** (0.03) \\
log(Read Duration) & -0.06** (0.02) & -0.14** (0.04) & -0.05** (0.011) & -0.13** (0.03) & -0.08** (0.013) & -0.18** (0.03) \\
Total Read Duration (min) & -13.86** (3.46) & -31.01** (7.70) & -13.66** (2.35) & -38.16** (6.61) & -13.88** (2.68) & -32.41** (6.27) \\
Time to Reply from Read Start Time (min) & -21.82 (11.48) & -48.88 (25.73) & -14.18 (7.90) & -39.63 (22.11) & -13.70 (8.03) & -32.09 (18.78) \\
Total Session Duration (min) & -2.05 (3.78) & -4.40 (8.14) & 3.74 (2.39) & 10.15 (6.47) & 7.89* (3.16) & 17.74* (7.09) \\
Total Scheduled Meetings Attended & -0.12 (0.15) & -0.25 (0.33) & 0.16 (0.10) & 0.44 (0.26) & 0.36** (0.13) & 0.80** (0.30) \\
Joined Scheduled Meeting 5 Min Late & 0.05 (0.03) & 0.10 (0.07) & 0.03 (0.03) & 0.09 (0.07) & 0.06 (0.03) & 0.13 (0.07) \\
Left Scheduled Meeting 10 Min Early & 0.03 (0.03) & 0.06 (0.07) & 0.03 (0.03) & 0.09 (0.07) & 0.09** (0.03) & 0.20** (0.07) \\
Wrd, Exl, Ppt Files Edited & 0.24** (0.09) & 0.53** (0.19) & 0.02 (0.07) & 0.05 (0.20) & 0.12 (0.06) & 0.26 (0.14) \\
Word Docs Created & 0.06** (0.02) & 0.12** (0.05) & 0.02 (0.02) & 0.06 (0.05) & 0.010 (0.015) & 0.02 (0.03) \\
Primary Editor Word Docs & 0.07** (0.02) & 0.14** (0.05) & 0.03 (0.02) & 0.08 (0.05) & 0.009 (0.015) & 0.02 (0.03) \\
Read Others' Word Docs & 0.11** (0.03) & 0.23** (0.05) & 0.003 (0.02) & 0.008 (0.05) & 0.05** (0.02) & 0.10** (0.04) \\
Time to Complete (days) & -1.07* (0.47) & -1.81* (0.79) & -0.10 (0.38) & -0.26 (0.97) & -1.52** (0.40) & -2.74** (0.73) \\
Completed Word Docs in $\le$ 1 wk & 0.04* (0.02) & 0.08* (0.04) & 0.02 (0.013) & 0.06 (0.04) & 0.000 (0.012) & 0.000 (0.03) \\
Completed Word Docs in $>$ 1 wk & 0.007 (0.008) & 0.02 (0.02) & 0.006 (0.005) & 0.02 (0.015) & -0.011* (0.005) & -0.02* (0.011) \\
\hline
\multicolumn{7}{l}{\footnotesize * $p < 0.05$, ** $p < 0.01$} \\
\end{tabular}%
}
\caption{Effect of Copilot on All Metrics by Industry Groups}
\label{table:copilot_by_industry_groups}
\end{table}

\end{landscape}
\end{document}